\documentstyle [12pt]{article}

\begin{document}
\begin{titlepage}
\begin{center}
{\Large\bf Analytic regularization of the 
Yukawa Model at Finite Temperature}\\

\vspace{.3in}{\large\em A.P.C.Malbouisson}\\
 Centro Brasileiro de Pesquisas Fisicas-CBPF\\
 Rua Dr.Xavier Sigaud 150, Rio de Janeiro, RJ 22290-180 Brazil

\vspace{.3in}{\large\em B.F. Svaiter}\\

Instituto de Matematica Pura e Aplicada-IMPA\\ Estrada Dona Castorina 110,
Rio de Janeiro 22460, RJ, Brazil\\
\vspace{.2in}

\vspace{.3in}{\large\em N.F. Svaiter*}\\
 Centro Brasileiro de Pesquisas Fisicas-CBPF\\
 Rua Dr.Xavier Sigaud 150, Rio de Janeiro, RJ 22290-180 Brazil 
\subsection*{\\Abstract}
\end{center}

We analyse the one-loop 
fermionic contribution for the scalar effective potential 
in the temperature dependent Yukawa model. In order to regularize 
the model a mix between dimensional and analytic regularization 
procedures is used. We find a general expression for the 
fermionic contribution in arbitrary spacetime dimension. It is found 
that in $D=3$ this contribution is finite.

\nopagebreak
Pacs number: 11.10.Ef, 11.10.Gh. 
\\* e-mail: nfuxvai@lca1.drp.cbpf.br

\end{titlepage}
\newpage\baselineskip .37in
\section {Introduction}\

Recently there has been much interest in the phase structure of theories 
involving scalar fields
presenting  spontaneous symmetry breaking.  
Many aplications have been done in the Weinberg-Salam model and in 
grand-unified theories. The temperature generally is the parameter 
whose variation induces  the transition from the broken to the 
unbroken phase, at least for the most current systems that  
develop first 
or second order phase transitions. 

To describe a second order phase 
transition the variation of the mass with the temperature is the most 
important fact. On the other hand the dependence of the coupling 
constant with the temperature may induce a first order phase transition 
in the scalar sector, as suggested by two of the authors in a recent work
\cite{Adolfo1}.

We start from the Yukawa model and we analyse 
the contribution  coming 
from the fermionic loops
for the temperature dependent scalar effective potential. The ultraviolet 
divergences are dealed with the method 
of analytic regularization \cite{Bol}.
We recall that
the basic idea of this technique is to replace the denominator of the 
propagator $(p^{2}-m^{2}+i\epsilon)$ by $(p^{2}-m^{2}+i\epsilon)^
{1+\alpha}$ where $\alpha$ is the regulating parameter initially 
taken to be 
large enough. Consequently in a open connected set of points in the 
complex plane $\alpha$ the Feynman amplitudes are analytic. Then it is 
possible to analytically continue the Feynman expressions to the whole 
complex plane. In the Laurent expansions of these expressions we 
can identify the counterterms as the polar terms in the 
analytic extensions at some points of the complex plane. 

To deal with finite temperature field theory using the imaginary 
time formalism we will have to use dimendional regularization in the 
momenta and deal with the Matsubara sums using another method. The 
most popular method to deal with the Matsubara sum is an analytic 
extension away from the discrete complex energies down to the real 
axis, with the replacement of the energies sums by countour integrals 
\cite{Kapusta}. If we are interested in systems at high temperature the 
decoupling theorem \cite{Carrazone} allow us to use the dimension reduction 
method $(DR)$.  This approach has been used by many authors \cite{Appelquist}.
The basic idea is that in the imaginary time formalism the free propagator 
has a form  $(\omega_{n}^{2}+p^{2}+m^{2})^{-1}$. The Matsubara 
frequency act like a mass so in the high temperature regime the non-static 
$(n \neq 0)$ modes decouple, and we have a three dimensional theory. 
In other words, the only modes whose contribution do not fall of 
exponentially at distances much greater than $\beta$ are the $(n=0)$ modes 
of the bosons. Integration over the fermionic modes and the non-zero modes 
of the bosons result in a three dimensional theory. Of course this effective 
model will describe the original model only for distances $R>>\beta$.
As was stressed by Landsman \cite{Landsman} the standard summation method
\cite{Kapusta} based on analytic continuation do not work in the 
dimensional reduction approach. Instead, we have first to compute momentum 
integrals using dimensional regularization and to deal with the Matsubara 
sums, a inhomogeneous zeta function analytic regularization has to 
be performed.

Recently
such technique has been used to study different models at finite 
temperature. Ford and Svaiter \cite{Ford} and Malbouisson and Svaiter \cite{Adolfo1}
\cite{Adolfo2} studied the  $\lambda\varphi^{4}$ and the 
Efimov-Fradkin (truncated or not) model at finite temperature. 
The possibility 
of  vanishing the temperature dependent coupling 
constants in these models has been investigated. 
In the first paper,  assuming a 
non-simply connected spatial section, the thermal 
and topological contributions 
to the renormalized mass and coupling constant in the 
$(\lambda\varphi^{4})_{D=4}$ model 
was obtained at the one-loop approximation. 
In the second one, the authors extend 
the discussion of the massive self-interacting $\lambda\varphi^{4}$ model 
to an arbitrary D-dimensional spacetime with trivial topology 
of the 
spacelike sections. The main result is that the possibility of  
a first order phase transition drived by the temperature 
dependent coupling constant, 
in the region where the model is super-renormalizable arises.
The discussion in the case of 
a scalar model with non-polynomial interaction Lagrange density 
(the Efimov-Fradkin model) 
has been done in the third work. For $D>2$ it was proved 
that at least 
two coupling constants of the truncated model may vanish and 
become negative 
by effect of temperature changes, while in the non-truncated model 
all the coupling 
constants remain positive for any 
temperature. The method used in the above quoted papers 
could provide an almost natural way
to investigate stability regimes in finite temperature QFT models.

It has been often sugested that the thermal contributions 
to the renormalized coupling constants of quantum models may 
bring up non-trivial effects. For instance, 
Gross, Pisarski and Yaffe \cite{Gross}
argue that in  finite temperature $(QCD)_{4}$ the effective 
coupling constant $g(\Lambda)$ 
decreases as the temperature or density is raised. 
In fact, they show in a perturbative context that 
at the first non-trivial 
order $(QCD)_{4}$ should be asymptotically free at high temperature 
or pressure. In this approximation
it is expected that at high temperatures thermal excitations 
produce a plasma of quarks and gluons which screen all (color) eletric 
flux. Such a transition from a low temperature confined phase to 
a high temperature color screening phase has been also 
investigated by Polyakov
\cite{Polyakov} and Susskind \cite{Susskind} and others in lattice 
gauge theories. Such results have important astrophysical 
applications in the study of neutron stars or primeval universe models.

The main goal of this paper is to investigate the 
one-loop femionic contribution to the scalar effective potential
at finite
temperature assuming that 
bosons and fermions interact via a Yukawa coupling. 
The outline of the paper is the following: in section II we briefly 
review  the formalism of the 
effective potential. In section III the fermionic contribution to 
the effective potential is obtained. In section IV the singularity 
structure of the one-loop fermionic contribution to the scalar effective 
potential is studied
Conclusions are given in section V. In
this paper we use $\hbar=k_{B}=c=1$.

\section{ The effective action and the effective potential at
 zero temperature.}

In this section we will briefly review the basic features of the 
effective potential associated with a
real massive self-interacting scalar field at zero temperature. 
Although the formalism of this 
section may be found in standard texbooks, we recall here its 
main results for 
completeness. Let us consider a real 
massive scalar field $\varphi(x)$ with the usual 
$\lambda\varphi^{4}(x)$ self-interaction, defined in a static
spacetime. Since the manifold is static, there is a global timelike 
Killing vector field orthogonal to the spacelike sections. 
Due to this fact,
energy and thermal equilibrium have a precise meaning. For 
the sake of simplicity, 
let us suppose that the manifold is flat. In the path 
integral approach, the
basic object is the generating functional,
$$
Z[J]= < 0,out | 0,in >=
$$
\begin{equation}
\int{\cal D}[\varphi]\exp \{i[S[\varphi]+
\int d^{4}x J(x)\varphi(x)]\}
\end{equation}
where $ {\cal D}[\varphi]$ is an apropriate
 integration measure and $ S[\varphi]$ is the classical
action associated with the scalar field. The 
quantity $ Z[J] $ gives the
transition amplitude from the initial 
vacuum $ |0, in > $ to the final vacuum
$ |0,out > $ in the presence of some 
source $ J(x)$, which is zero outside
 some interval $ [-T,T] $ and inside this interval 
is switched adiabatically
 on and off. Since we are 
interested in the connected part of the time ordered products
 of the fields, 
we take the connected generating functional $W[J]$, 
as usual. This quantity is
defined in terms of the vacuum persistent amplitude by

\begin{equation}
e^{i W[J]}= Z[J],
\end{equation}
and the connected $n$-point functions $ G^{(n)}_{c}(x_{1},x_{2},..,x_{n})$ 
are

\begin{equation}
G^{(n)}_{c}(x_{1},x_{2},..,x_{n})=\frac{\delta^{n} W[J]}{\delta J(x_{1})...
\delta J(x_{n})}|_{J=0}.
\end{equation}

Expanding $ W[J]$ in a functional Taylor series, 
the n-order coefficient of this
series will be the sum of all connected Feynman 
diagrams with $n$ external
legs, i.e. the connected Green's functions defined by eq.(3). Then
\begin{equation}
W[J]=\sum^{\infty}_{n=0}\frac{1}{n!}\int 
d^{4}x_{1}..d^{4}x_{n}~ G^{(n)}_{c}(x_{1},x_{2}..
..x_{n}) J(x_{1})J(x_{2})..J(x_{n}).
\end{equation}

The classical field $\varphi_{0}(x)$ is given by the normalized vacuum 
expectation value of the field
\begin{equation}
\varphi_{0}(x)=\frac{\delta W}{\delta J(x)}=
\frac{<0, out|\varphi(x)|0, in >_{J}}{<0, out|0, in >_{J}},
\end{equation}
and the effective action $\Gamma[\varphi_{0}]$ is 
obtained by performing a 
functional Legendre transformation
\begin{equation}
\Gamma[\varphi_{0}]=W[J]-\int d^{4}x J(x)\varphi_{0}(x).
\end{equation}

Using the functional chain rule and the definition 
of $\varphi_{0}$ given
by eq.(5) we have
\begin{equation}
\frac{\delta\Gamma[\varphi_{0}]}{\delta\varphi_{0}}=-J(x).
\end{equation}

Just as $W[J]$ generates the connected Green's functions by means of a 
functional
Taylor expansion, the effective action can be 
represented as a functional 
power series around the value $\varphi_{0}=0$, 
where the coeficients are 
just the proper $n$-point
functions $ \Gamma^{(n)}(x_{1},x_{2},..,x_{n})$ i.e.,
\begin{equation}
\Gamma[\varphi_{0}]=
\sum^{\infty}_{n=0}\frac{1}{n!}\int d^{4}x_{1}d^{4}x_{2}..
.d^{4}x_{n}~\Gamma^{(n)}(x_{1},x_{2},..,x_{n})
~\varphi_{0}(x_{1})\varphi_{0}(x
_{2})..\varphi_{0}(x_{n}).
\end{equation}

The coefficients of the above functional expansion 
are expressed in terms of the connected 
one-particle irreducible diagrams $(1PI)$. Actually, 
$\Gamma^{(n)}(x_{1},x_{2},..
.,x_{n})$ is the sum of all $ 1PI$ Feynman diagrams 
with $n$ external legs.
 Writing the effective action in powers of momentum 
(around the point where
all external momenta vanish) we have
\begin{equation}
\Gamma[\varphi_{0}]=\int d ^{4}x\biggl(-V(\varphi_{0})+
\frac{1}{2}(\partial_{\mu}
\varphi)^{2} Z[\varphi_{0}]~+~..\biggr).
\end{equation}

The term $ V(\varphi_{0})$ is called the effective 
potential \cite{Coleman} .To express 
$ V(\varphi_{0})$ in terms of the $1PI$ 
Green's functions, we write
$ \Gamma^{(n)}(x_{1},x_{2},..,x_{n})$ in 
momentum space,
\begin{equation}
\Gamma^{(n)}(x_{1},x_{2},..,x_{n})=
\frac{1}{(2\pi)^{n}}\int d^{4}k_{1}d^{4}k_{2}..d^{4}k_{n} (2\pi)^{4}
\delta(k_{1}+ k_{2} +..k_{n})\ e^{i(k_{1}x_{1}+..k_{n}x_{n})}
\tilde \Gamma^{(n)}(
x_{1},x_{2},..,x_{n}).
\end{equation}

Assuming that the model is translationally 
invariant, i.e. $\varphi_{0}$  
is constant over the manifold, we have
\begin{equation}
\Gamma[\varphi_{0}]=\int d^{4}x \sum^{\infty}_{n=1}
\frac{1}{n!}\biggr(\tilde\Gamma^{(n)}(0,0,..
.)(\varphi_{0})^{n}+...\biggr).
\end{equation}
If we compare eq.(9) with eq.(11) we obtain
\begin{equation}
V(\varphi_{0})= -\sum_{n}\frac{1}{n!}\tilde\Gamma^{(n)}
(0,0,..)(\varphi_{0})^{n},
\end{equation}
then $ \frac{d^{n}V}{d\varphi^{n}_{0}} $ is the 
sum of the all $1PI$ diagrams
carring zero external momenta. Assuming that the 
fields are in equilibrium 
with a thermal reservoir at temperature 
$\beta^{-1}$, in the Euclidean 
time formalism, the effective potential 
$V(\beta,\varphi_{0})$ can be 
identified with the free energy density and 
can be calculated by 
imposing periodic (antiperiodic) boundary 
conditions on the bosonic (fermionic) fields.

\section{ The one-loop effective potential of the Yukawa model at
 zero and finite temperature.}\

Let us consider a system consisting of bosons and fermions fields 
interacting via a Yukawa coupling in thermal equilibrium with a 
reservoir at temperature $\beta^{-1}$.
They are defined on 
a four dimensional flat spacetime with trivial topology 
of the spacelike sections. In the zero temperature case
the generating functional for the scalar and fermionic fields 
correlation functions is given 
by:

\begin{equation}
Z[\eta,\bar{\eta},J]=
\int{\cal D}\psi{\cal D}\bar{\psi}{\cal D}\varphi
\exp\{i[S[\bar{\psi},\psi,\varphi]+
\int d^{4}x \bar{\psi}\eta+\bar{\eta}\psi+J\varphi]\}
\end{equation}
where $\bar{\psi(x)}$, $\psi(x)$, $\bar{\eta(x)}$ and $\eta(x)$ are 
elements of the Grassmann algebra and $\varphi(x)$ and $J(x)$ are 
commuting variables.

 The perturbativelly renormalizable action has the form,

\begin{equation}
S[\bar{\psi},\psi,\varphi]=\int d^{4}x
\left(\frac{1}{2}(\partial_{\mu}\varphi_{b})^{2}-
\frac{1}{2}m_{0}^{2}\varphi_{b}^{2}+V(\varphi_{b})+
\bar{\psi}_{b}(i\not{\partial}-M_{0}-g_{0}\varphi_{b})\psi_{b}\right)
\end{equation}
where $V(\varphi_{b})=\frac{\lambda}{4!}\varphi_{b}^{4}$ and 
$m_{0}$, $M_{0}$ are respectivelly the boson and the fermion 
bare masses and $\lambda_{0}$ and $g_{0}$ are the bare 
coupling constants. Of course  $\varphi_{b}$ and $\psi_{b}$
are bare bosonic and fermionic field.

The most general divergent terms are of the type \cite{Zinn} 

\begin{equation}
\begin{array}{c}
-\Gamma^{div}=\int d^{4}x
\left( \frac{1}{2}\delta Z_{\varphi}(\partial_{\mu}\varphi)
^{2}-\frac{1}{2}\delta m^{2}\varphi^{2}+\right.\\
\left.+\delta Z_{\psi}i\bar{\psi}(i\not{\partial}\psi-\delta M\bar{\psi}\psi-
g\delta Z_{g}\bar{\psi}\psi\varphi+\frac{1}{4}\delta\lambda\varphi^{4}+
\frac{1}{3}\delta\sigma\varphi^{3}+\delta c\varphi\right)
\end{array}
\end{equation}

Although the action is renormalizable, the model is not 
multiplicatively renormalizable. To circumvect 
this difficulty and to allow the 
theory to become multiplicatively renormalizable we shall introduce at the 
tree level action all terms which we expect to be generated by the 
renormalization procedure,
i.e.

\begin{equation}
S(\bar{\psi},\psi,\varphi)=
\int d^{4}x\left(\frac{1}{2}\partial_{\mu}\varphi\partial^{\mu}\varphi-
\sum_{n=1}^{4}\lambda_{n}\varphi^{n}+ 
\bar{\psi}(i\not{\partial}-M-g\varphi)\psi\right)+counterterms,
\end{equation}

where $\lambda_{2}=\frac{1}{2}m^{2}$, $\lambda_{3}=\frac{\sigma}{3!}$ 
and $\lambda_{4}=\frac{\lambda}{4!}$.

As usual, perturbation theory is generated by,

\begin{equation}
Z(\bar{\eta},\eta,J)=
exp\{-i\int d^{4}x ( i^{3}g\frac{\delta^{3}}{\delta\eta\delta\bar{\eta}\delta J}
+V(\frac{\delta}{\delta J}))\}Z_{0}(\bar{\eta},\eta,J)
\end{equation}
where 
\begin{equation}
Z_{0}(\bar{\eta},\eta,J)=
exp{-i\int d^{4}x d^{4}y (\bar{\eta}(x)\Delta_{F}(x-y)\eta(y)+
\frac{1}{2}J(x)\Delta(x-y)J(y))}
\end{equation}
with $\Delta_{F}(x-y)$ and $\Delta(x-y)$ being respectively 
the fermionic and bosonic propagator functions,

\begin{equation}
\Delta_{F}(x-y)=(i\not{\partial}_{x}+M)\Delta(x-y),
\end{equation}
and 

\begin{equation}
\Delta(x-y)=
\frac{1}{(2\pi)^{4}}\int d^{4}p\frac{e^{-ip(x-y)}}{p^{2}-m^{2}+i\epsilon}.
\end{equation}

From the above formulas, following a porcedure entirely analogous 
to that described in the preceding section for the pure scalar case
it is easy to get 
the fermionic contribution to the effective potential $V(\varphi_{0})$

\begin{equation}
V(\varphi_{0})\int d^{4}x=
i ln\int^{0}_{0} \bar{d\psi}d\psi~ exp\{i\int d^{4}x 
\bar{\psi}(i\not{\partial}-M-g\varphi)\psi\}
\end{equation}

After a Wick rotation to Euclidean space
and using the rules for
Grassmann integrals we get the contribution from the single 
fermionic loops to the 
scalar effective potential

\begin{equation}
V(\varphi_{0})\int d^{4}x=-ln\,det(i\not{\partial}_{E}-M-g\varphi_{0})
\end{equation}

Using a well known result 

\begin{equation}
log\,det(M+g\varphi_{0})=tr\,log(M+g\varphi_{0}),
\end{equation}
we have,
 
\begin{equation}
ln\, det(i\not{\partial}_{E}-M-g\varphi_{0})=tr\,log(i\not{\partial}_{E})-
\sum^{\infty}_{s=1}\frac{(-i)^{s}}{s}(M+g\varphi_{0})^{s}
tr(\frac{1}{\not{\partial}_{E}})^{s}.
\end{equation}

Using a Fourier representation for $\frac{1}{\not{\partial}_{E}}$ and 
taking into account that the contributions from odd values of $s$ 
in the above sum vanish, 
it is possible to recast the fermionic contribution to the effective 
potential in the form \cite{Swanson},

\begin{equation}
V(\varphi_{0})=
4\sum^{\infty}_{s=1}\int\frac{d^{4}p}{(2\pi)^{4}}
\frac{(-1)^{s}}{2s}\frac{(M+g\varphi_{0})^{2s}}{(p^{2}_{E})^{s}}.
\end{equation}

In the finite temperature case using the Matsubara formalism we have to
perform the replacements 
$\omega\rightarrow
\omega_{n}=\frac{2\pi}{\beta}(n+\frac{1}{2})$
and $\frac{1}{2\pi}\int dq^{0}_{E}=\frac{1}{\beta}\sum_{n}$. Then the 
contribution from the single fermionic 
loops to the effective potential is given by

\begin{equation}
V(\varphi_{0},\beta)=
\frac{2}{\beta}\sum^{\infty}_{s=1}\sum^{+\infty}_{n=-\infty}
\int\frac{d^{d}p}{(2\pi)^{d}}
\frac{(-1)^{s}}{s}\frac{(M+g\varphi_{0})^{2s}}{(\omega^{2}_{n}+q^{2})^{s}}.
\end{equation}

Let us define the quantities,

\begin{equation}
a=(\frac{1}{\beta\mu})^{2}
\end{equation}

\begin{equation}
\phi=\frac{\varphi_{0}}{2\pi\mu}
\end{equation}
and
\begin{equation}
\gamma=\frac{M}{2\pi\mu},
\end{equation}
where $\mu$ is a parameter with mass dimension
introduced to deal with dimensionless quantities performing 
analytic extensions. First we use dimensional regularization 
going to a generic $D$-dimensional spacetime. Then eq.(26) becomes 

\begin{equation}
V(\phi,\beta)=\mu^{D}\sum^{\infty}_{s=1}
a^{\frac{D}{2}-s}f(D,s)(\gamma+g\phi)^{2s}
\sum^{\infty}_{n=-\infty}\frac{1}
{((n+\frac{1}{2})^{2})^{s-\frac{d}{2}}}.
\end{equation}
where $f(D,s)$ is given by:

\begin{equation}
f(D,s)=\frac{2\pi^{\frac{d}{2}}}{\Gamma(s)}\Gamma(s-\frac{d}{2})
\frac{(-1)^{s}}{(2\pi)^{2s}}.
\end{equation}

Before going one some commments are in order. It is well 
known \cite{Leibrant} that dimensional regularization techniques 
for massless fields can not led to definite results due to the 
presence  of infrared divergences \cite{Leibrant}. Since we are regularizing 
only a $d=D-1$ dimensional integral, this procedure is equivalent 
to inserting a mass into the $d$ dimensional integral. In other 
words, the Matsubara frequencies play the role of "masses" in the 
integral provided that we exclude the limit $\beta\rightarrow \infty$
which means that we {\it must restrict ourselves 
to non-zero temperatures}. Another point is that in order to evaluate 
the one-loop finite temperature diagrams the usual approach is to 
express the integrand as a countour integral \cite{Kapusta}. In this 
paper we use another technique still aplying the principle of the analytic 
extension.

In the next section we will analyse the singularity 
structure of the inhomogeneous Riemann zeta function 
and other factors appearing in eq.(30)
in order to identify the divergent terms in the 
fermionic contribution to the effective potential. We 
start by analytically regularizing the model.

\section{ The singularity structure of the fermionic 
contribution to the effective potential.}\

As we remarked before the fermionic contribution to the 
effective potential is ill defined due to the singularities of the 
gamma function that appears in $f(D,s)$ and the singularities 
in the Matsubara sum. The Matsubara sum may be expressed in terms of the  generalized inhomogeneous Riemann zeta function, which 
can be analytically extended to a meromorphic 
function in the whole complex $s$ plane. The polar terms 
must be removed in the renormalization procedure. In order 
to identify these poles let us first recall the definition of the 
inhomogeneous 
Riemann zeta function or Hurwitz zeta function \cite{Gradstein}

\begin{equation}
\zeta(z,q)=
\sum^{\infty}_{n=0}\frac{1}{(n+q)^{z}},
\end{equation}
which is analytic for $Re(z)>1$. 

After some manipulations it is possible to
express the Matsubara sum in eq.(30) in terms of $\zeta(z,q)$
and write $V(\phi,\beta)$ in the 
form,

\begin{equation}
V(\phi,\beta)=2\mu^{D}\sum^{\infty}_{s=1}
a^{\frac{D}{2}-s}f(D,s)(\gamma+g\phi)^{2s}
\zeta(2s-d,\frac{1}{2}).
\end{equation}

To analytically extend the inhomogeneous 
Riemann zeta function, we go along the 
following steps:
first using the Euler representation for the Gamma function we write
it as 

\begin{equation}
\zeta(z,q)=
\frac{1}{\Gamma(z)}\int^{\infty}_{0}
dt ~t^{z-1}\frac{e^{\frac{-t}{2}}}{1-e^{-t}}.
\end{equation}

Next, we split the integral from zero to infinity in two integrals, 
from zero to one and from one to infinity. The second one is 
an analytic function of $z$, the divergences being associated to the zero 
limit of the first integral. Then using a Bernoulli representation 
for the integrand it is possible to get the following expression to 
the analytic extension of $\zeta(z,\frac{1}{2})$

\begin{equation}
\zeta(z,\frac{1}{2})=g_{1}(z)+\frac{1}{\Gamma(z)}\sum_{n=0}^{\infty}
\frac{B_{n}(\frac{1}{2})}{n!}\frac{1}{z+n-1}
\end{equation}

where $g_{1}(z)$ is given by

\begin{equation}
g_{1}(z)= \frac{1}{\Gamma(z)}\int^{\infty}_{1}dt\, t^{z-1}
\frac{e^{\frac{t}{2}}}
{e^{t}-1},
\end{equation}

and the $B_{n}(x)$ are the Bernoulli coefficients \cite{Gradstein}.
We remark that in the literature there is another 
formula for the analytic extension of the inhomogeneous 
Riemann zeta function; the Hermite formula \cite{Hermite} given by

\begin{equation}
\zeta(z,q)=\frac{1}{2q^{z}}+\frac{q^{1-z}}{z-1}+2\int^{\infty}_{0}
(q^{2}+y^{2})^{\frac{-z}{2}}\sin(z\arctan\frac{y}{q})\frac{1}{e^{2\pi y}-1}dy.
\end{equation}

Of course the analytic extension must be uniquely 
defined and these are only 
different representation of the same analytic extension. 
Substituting the analytic extension given 
by eq.(35) in the 
fermionic contribution to the effective potential $V(\phi,\beta)$ 
we get,

\begin{equation} 
\begin{array}{lcr}
V(\phi,\beta)&=&\mu^{D}\sum^{\infty}_{s=1}
a^{\frac{D}{2}-s}h(D,s)(\gamma+g\phi)^{2s}
\frac{1}{\Gamma(
-\frac{D}{2}+s+1)}\\[1em]
&&\quad\left(\int^{\infty}_{1}dt\, t^{2s-D}\frac{e^{\frac{t}{2}}}
{e^{t}-1}+
\sum_{n=0}^{\infty}
\frac{B_{n}(\frac{1}{2})}{n!}\frac{1}{2s-D+n}\right).
\end{array}
\end{equation}
where the regular function $h(D,s)$ is given by,
\begin{equation}
h(D,s)=2\frac{(-1)^{s}}{s}\frac{(2\pi^{\frac{1}{2}})^{D-4s}}{\Gamma(s)}.
\end{equation}

Let us analyse the two cases $D=3$ and $D=4$ separately.
For the case $D=3$ we have
\begin{equation}
\begin{array}{c}
V(\phi,\beta)=\mu^{3}\sum^{\infty}_{s=1}
a^{\frac{3}{2}-s}h(3,s)(\gamma+g\phi)^{2s}
\frac{2}{\Gamma(
-\frac{3}{2}+s+1)}\\[1em]
\left(\int^{\infty}_{1}dt\, t^{2s-3}\frac{e^{\frac{t}{2}}}
{e^{t}-1}+
\sum_{n=0}^{\infty}
\frac{B_{n}(\frac{1}{2})}{n!}\frac{1}{2s-3+n}\right)
\end{array}
\end{equation}

The fermionic contibution to the effective potential is finite. There 
is no ultraviolet divergences in $D=3$. One would not normally 
expect this since the tadpole graph is ultraviolet divergent $(s=1)$.
This situation is very similar to the calculation of the 
renormalized vacuum energy of scalar fields confined in boxes (Casimir 
energy) \cite{Benar}. Dolan and Nash used the zeta 
function analytic regularization method to obtain the Casimir energy 
of conformally coupled scalar field confined in odd and 
even dimensional spheres \cite{Dolan}. 
They obtained that for odd dimensional spheres 
(even space-time dimension) there is a pole in the 
point of interested, being necessary the introduction of a counterterm,
while for even dimensional spheres (odd dimensional space-time) the result
obtained is naturally finite. No renormalization is needed.
For a careful study of this subject see ref.\cite{Blau}.

For the case $D=4$ we have 
\begin{equation} 
\begin{array}{rcl}
V(\phi,\beta)&=&\mu^{4}\sum^{\infty}_{s=1}
a^{2-s}h(4,s)(\gamma+g\phi)^{2s}
\frac{1}{\Gamma(
s-1)}\\[1em]
&&\left(\int^{\infty}_{1}dt\, t^{2s-4}\frac{e^{\frac{t}{2}}}
{e^{t}-1}+
\frac{B_{0}(\frac{1}{2})}{2s-4}+\frac{B_{2}(\frac{1}{2})}{4s-4}
+\sum_{n=3}^{\infty}
\frac{B_{n}(\frac{1}{2})}{n!}\frac{1}{2s-4+n}\right).
\end{array}
\end{equation}

Note that the factor $\Gamma^{-1}(s-1)$ just cancels 
the pole from the term $n=2$ in the sum over $n$. The 
pole comming from the term $n=0$ in the sum must be 
canceled by the introduction of a suitable counterterm. 
All other terms $s\ge 3$ are finite.

\section{Conclusion}

 The aim of this paper is to discuss an alternative method to deal 
with the Matsubara sum in a finite temperature field theory with bosons and 
fermions in interaction. We use this method to calculate the 
one-loop fermionic contribution to the scalar effective potential assuming 
the Yukawa coupling between fermions and bosons. Note that we 
are using a BPHZ scheme with subtraction at zero momentum of the 
Feynman integrals. Matsumoto, Ojima and Umezawa \cite{Ojima} claims that 
tha Matsubara method seems to produce temperature dependent divergences which 
disapear only after a summation over the Matsubara sums. We showed that 
the countertems are temperature independents. 

A curious observation is in order. We note that eq.(38) does not 
contains singularities for {\it any} odd space time dimension $D$, 
due to the fact that the sum is over integer values of $s$, and the 
Bernouilli coefficients $B_{n}(\frac{1}{2})=0$ for $n$ odd. For even
values of $D$ the fermionic contribution to the effective potential 
(see eq.(38)) has only a divergence due to the term $s=\frac{D}{2}$,
$n=0$.

It would 
be interesting to generalize the method if we consider that there is 
a non-zero fermion density \cite{Wolf}. This can be done 
introduzing a chemical potential $\sigma$. At finite temperature the chemical 
potential will change the Matsubara frequencies by
$\omega_{n}\rightarrow \omega_{n}+i\sigma$ \cite{Dashen}. In this case 
we have to  analytically extend the inhomogeneous Epstein zeta function
$\zeta(z,q)$ for complex $q$.  
This subject is under invetigation.

\section{Acknowlegement}

 This paper was supported by 
Conselho Nacional de Desenvolvimento Cientifico e Tecnologico do Brazil
(CNPq).

\end {document}